\documentclass{ptephy_v1}

\usepackage{amsmath} 
\usepackage{amsthm} 

\theoremstyle{plain}

\newcommand{\nn}{\nonumber \\}

\begin{document}

\title{Attractive gravity probe surface, positivity of quasi-local mass and Arnowitt-Deser-Misner mass expression}



\author{Tetsuya Shiromizu${}^{1,2}$}
\author{Keisuke Izumi${}^{2,1}$}
\affil{${}^1$Department of Mathematics, Nagoya University, Nagoya 464-8602, Japan}
\affil{${}^2$Kobayashi-Maskawa Institute, Nagoya University, Nagoya 464-8602, Japan}


\begin{abstract}%
Under certain conditions, it is shown that the positivity of the Geroch/Hawking quasi-local mass holds for 
the attractive gravity probe surfaces in any higher dimensions than three. 
We also comment on the Arnowitt-Deser-Misner mass. 
\end{abstract}

\subjectindex{xxxx, xxx}

\maketitle


\section{Introduction}

Recently the attractive gravity probe surface (AGPS)/loosely trapped surface have been proposed as indicators for 
the existence of the attractive (strong) gravity \cite{Shiromizu2017, Izumi2021, Izumi2023} and the areal inequalities 
have been proven under some assumptions as the Penrose inequality \cite{Penrose}. 
For the Schwarzschild spacetime, it is easy to see that the presence of the AGPS requires the positivity of 
the mass. In this paper, we will show that, under certain conditions, the Geroch/Hawking mass quasi-locally 
defined \cite{Geroch1973, Hawking1968} is positive on the AGPS for general cases. We will also point out 
that the existence of the sequence of the AGPSs directly show us the positivity of the Arnowitt-Deser-Misner (ADM) mass 
in asymptotically flat spacetimes. While one may not be able to show the positivity of the ADM mass using the 
Geroch mass for higher dimensions than four 
because the Gauss-Bonnet theorem is crucial \cite{Geroch1973}, our argument for the positivity 
of the Geroch/Hawking mass works regardless of spacetime dimensions. 

The rest of this paper is organized as follows. In Sect. 2, we review certain expression for the ADM mass and a 
natural generalization of the Geroch mass in higher dimensions. In Sect. 3, we show the positivity of the 
Geroch mass on the AGPS.  
In Sect. 4, we shorty comment on the positivity of the Hawking mass. 
In Sect. 5, inspired by these arguments for the AGPS, we present a new expression for the ADM mass and, for the comparison, 
new expressions of the Geroch/Hawking mass in terms of the Weyl tensor. 
Finally, we give a summary in Sect. 6. In Appendix A we confirm that a certain expression of the ADM mass is equivalent to that of the Geroch mass. 
In Appendix B, we show that one can immediately see the positivity of the Geroch/Hawking mass for the refined version of the 
AGPS introduced in Ref. \cite{Izumi2023}. 
We also discuss the lower bound and positivity of the Bartonik mass through the discussion 
in Ref. \cite{Izumi2023}. 

\section{ADM mass and Geroch mass in higher dimensions}

As Ref. \cite{Ashtekar1984}, it is known that the ADM mass in an asymptotically flat spacetime has the following form 
\footnote{Ref. \cite{Ashtekar1984} devoted to 
the four dimensional cases. For higher dimensions, see Ref. \cite{Tanabe2010}. In Appendix A, we give a quick check.}
\begin{eqnarray}
m_{\rm ADM} & = & -\frac{1}{8\pi (n-3)}\int_{S_\infty} r {}^{(n-1)}R_{ab}r^ar^bdA, \label{ADM-ricci}
\end{eqnarray}
where $S_\infty$ is the $(n-2)$-sphere at spatial infinity, ${}^{(n-1)}R_{ab}$ is the Ricci tensor of the 
$(n-1)$-dimensional asymptotically flat spacelike hypersurface $\Sigma$, $r^a$ is the outward unit normal vector to $S_\infty$ in $\Sigma$ and $r$ is 
the radial coordinate near the spatial infinity. Using the double traced Gauss equation, the ADM mass is also expressed by 
\footnote{For four dimensions, see Appendix in Ref. \cite{Shiromizu2023}.}
\begin{eqnarray}
m_{\rm ADM}
& = & \frac{1}{16\pi (n-3)}\int_{S_\infty} r \left( -{}^{(n-1)}R+{}^{(n-2)}R+\tilde k_{ab} \tilde k^{ab}-\frac{n-3}{n-2} k^2 \right)dA \nonumber \\
& = & \frac{1}{16\pi (n-3)}\int_{S_\infty} r \left({}^{(n-2)}R-\frac{n-3}{n-2} k^2 \right)dA, \label{ADM2}
\end{eqnarray}
where, in the second equality, we used the fact that the $(n-1)$-dimensional Ricci scalar ${}^{(n-1)}R$, 
the traceless part of the extrinsic curvature of $S_\infty$, $\tilde k_{ab}$, rapidly decays near the spatial infinity. 
Here, ${}^{(n-2)}R$ is the Ricci scalar and $k$ is the trace of the extrinsic curvature of $S_\infty$ in $\Sigma$.  

In $n$-dimensions, one may define the Geroch mass on an $(n-2)$-dimensional compact surface $S$ in 
$\Sigma$ as \footnote {The current coefficient differs from that in 
Ref. \cite{Shiromizu2010} because the latter does not pay attention to the proportional coefficient.} \cite{Geroch1973, Shiromizu2010}
\begin{eqnarray}
m_G(S) =  \frac{A_{n-2}^{\frac{1}{n-2}}}{16\pi (n-3)\Omega_{n-2}^{\frac{1}{n-2}}}\int_{S} 
\left({}^{(n-2)}R-\frac{n-3}{n-2} k^2 \right)dA.\label{geroch1}
\end{eqnarray}
This may be attained through the replacement of $S_\infty$ and $r$ by $S$ and $(A_{n-2}/\Omega_{n-2})^{\frac{1}{n-2}}$ in Eq. (\ref{ADM2}) respectively, 
where $A_{n-2}$ is the area of $S$ and $\Omega_{n-2}$ is the volume element of the $(n-2)$-dimensional unit round sphere, that is, 
$\Omega_{n-2}=2\pi^{\frac{n-1}{2}}/\Gamma(\frac{n-2}{2})$. 

\section{Positivity of Geroch mass for AGPS}

Now we note that the following identity holds  
\begin{eqnarray}
r^aD_ak=-\frac{1}{2}k_{ab}k^{ab}-\frac{1}{2}k^2+\frac{1}{2}{}^{(n-2)}R-\frac{1}{2}{}^{(n-1)}R-\varphi^{-1}{\cal D}^2 \varphi,\label{rDk}
\end{eqnarray}
where $D_a$ and ${\cal D}_a$ are the covariant derivatives of $\Sigma$ and $S$, respectively. Here, $k_{ab}$ is the extrinsic curvature of 
$S$ in $\Sigma$ and $\varphi$ is the lapse function for the 
radial direction. Using this, we can rewrite the Geroch mass as 
\begin{eqnarray}
m_G(S) =  \frac{A_{n-2}^{\frac{1}{n-2}}}{8\pi (n-3)\Omega_{n-2}^{\frac{1}{n-2}}}\int_{S} 
\left[r^aD_ak+\frac{1}{n-2}k^2+\frac{1}{2}{}^{(n-1)}R+\frac{1}{2}\tilde k_{ab} \tilde k^{ab}+\varphi^{-2}({\cal D}\varphi)^2   \right]dA.
\label{geroch2}
\end{eqnarray}
Then we consider an attractive gravity probe surface (AGPS) defined as an $(n-2)$-dimensional 
compact surface satisfying two inequalities at each point on the surface \cite{Izumi2021, Izumi2023}
\begin{eqnarray}
&&r^aD_ak \geq \alpha k^2, \label{defagps} \\
&& k> 0,\label{defagps2}
\end{eqnarray}
where $\alpha$ is a constant satisfying $\alpha>-1/(n-2)$. 
If ${}^{(n-1)}R \geq 0$ holds on the AGPS,  
one can show that the Geroch mass is bounded from below as 
\footnote{Equation (\ref{defagps2}) is used in the current argument only to see that $\int_S k^2dA >0 $ holds, and is the condition that 
does not need to be taken into account unless $k=0$ for the entire $S$.}
\begin{eqnarray}
m_G(S) &\geq&  \frac{A_{n-2}^{\frac{1}{n-2}}}{8\pi (n-3)\Omega_{n-2}^{\frac{1}{n-2}}}\int_{S} 
\left[ \left(\alpha+\frac{1}{n-2} \right) k^2+\frac{1}{2}{}^{(n-1)}R+\frac{1}{2}\tilde k_{ab} \tilde k^{ab}+\varphi^{-2}({\cal D}\varphi)^2   \right]dA
\nn
&\geq& \frac{A_{n-2}^{\frac{1}{n-2}}}{8\pi (n-3)\Omega_{n-2}^{\frac{1}{n-2}}}  \left(\alpha+\frac{1}{n-2} \right) \int_{S} 
 k^2 dA
 \nn
&>&0\, .
\end{eqnarray}
Therefore, the conditions \eqref{defagps} and \eqref{defagps2} in the definition of the AGPS 
with the nonnegative scalar curvature ${}^{(n-1)}R \geq 0$ of $\Sigma$
guarantee the positivity of the Geroch mass. 
Note from the Hamiltonian constraint 
that ${}^{(n-1)}R \geq 0$ holds on the maximal spacelike hypersurface under 
the assumption that the spacetime geometry satisfies the Einstein equations with the dominant energy condition. 

It would be interesting to consider the limit case with $\alpha=-1/(n-2)$, 
although this case is out of the definition of AGPS.
Suppose that the scalar curvature is  nonnegative ${}^{(n-1)}R \geq 0$.
When $m_G$ vanishes, all inequalities of the above discussion should be equalities,
that is,
$r^aD_ak+k^2/(n-2)=0$, ${}^{(n-1)}R=0$, $\tilde k_{ab}=0$ and ${\cal D}_a \varphi=0$ hold on the surface.

\section{Positivity of Hawking mass for AGPS}

One may be also interested in the Hawking mass defined by 
\footnote{One may define the outgoing/ingoing null expansion rates by $\theta_\pm = (\kappa \pm k)/{\sqrt {2}}$.}
\begin{eqnarray}
m_H(S) =  \frac{A_{n-2}^{\frac{1}{n-2}}}{16\pi (n-3)\Omega_{n-2}^{\frac{1}{n-2}}}\int_{S} 
\left[{}^{(n-2)}R-\frac{n-3}{n-2} (k^2-\kappa^2) \right]dA, \label{hawking1}
\end{eqnarray}
where $\kappa=h^{ab}K_{ab}$, $h_{ab}$ is the induced metric of $S$ and $K_{ab}$ is the extrinsic curvature of 
spacelike hypersurface $\Sigma$. 
Since  the definitions of the Geroch/Hawking mass manifestly result in $m_H \geq m_G$, we see that $m_H$ of an AGPS is positive on 
$\Sigma$ with ${}^{(n-1)}R \geq 0$. 
Furthermore, under the assumption that the spacetime geometry satisfies the Einstein equations with the dominant energy condition,
we have the sharper lower bound for $m_H$. The Hamiltonian constraint gives
\begin{eqnarray}
{}^{(n-1)}R+\frac{n-3}{n-2}\kappa^2+2K_{(r)}\kappa-\tilde \kappa_{ab} \tilde \kappa^{ab}-2v_av^a=16\pi \rho,
\label{Hamcon}
\end{eqnarray}
where 
\begin{eqnarray}
K_{(r)}=K_{ab}r^ar^b,~\kappa_{ab}=h_a^ch_b^d K_{cd},~v_a=h_a^br^cK_{bc},
\end{eqnarray}
$\rho$ is the energy density and $\tilde \kappa_{ab}$ is the traceless part of $\kappa_{ab}$.
Using Eqs. (\ref{rDk}) and \eqref{Hamcon}, 
we have 
\begin{eqnarray}
m_H(S) & = &   \frac{A_{n-2}^{\frac{1}{n-2}}}{8\pi (n-3)\Omega_{n-2}^{\frac{1}{n-2}}}\int_{S} 
\Bigg[r^aD_ak+\frac{1}{n-2}k^2-K_{(r)}\kappa +8\pi \rho+ \frac{1}{2}\tilde k_{ab} \tilde k^{ab} \nonumber \\
& & +\frac{1}{2}\tilde \kappa_{ab} \tilde \kappa^{ab} +v_av^a+\varphi^{-2}({\cal D}\varphi)^2 \Bigg]dA. \label{hawking2} 
\end{eqnarray}
For the AGPS on the maximal slice ($K=K_{(r)}+\kappa=0$), we can prove the positivity of $m_H$ as  
\begin{eqnarray}
m_H(S) & \geq &   \frac{A_{n-2}^{\frac{1}{n-2}}}{8\pi (n-3)\Omega_{n-2}^{\frac{1}{n-2}}}\int_{S} 
\Bigg[ \left(\alpha+\frac{1}{n-2} \right)k^2+\kappa^2 +8\pi \rho+ \frac{1}{2}\tilde k_{ab} \tilde k^{ab} \nonumber \\
& & +\frac{1}{2}\tilde \kappa_{ab} \tilde \kappa^{ab} +v_av^a+\varphi^{-2}({\cal D}\varphi)^2 \Bigg]dA \nonumber \\
& > & 0.
\end{eqnarray}

As discussed for the Geroch mass, we consider the surface with $\alpha = -1/(n-2)$. 
When $m_H$ vanishes, $r^aD_ak+k^2/(n-2)=0$, $\rho=0$, $K_{ab}=0$, $\tilde k_{ab}=0$ and ${\cal D}_a \varphi=0$ hold on the surface.

\section{Asymptotic behavior and Weyl tensor expression}

In this section, inspired by the arguments in the previous two sections, we present a new expression for the ADM mass. 
In addition, since the ADM mass is written in terms of the Weyl tensor \cite{Ashtekar1984}, we 
will also rewrite the Geroch/Hawking mass in terms of the Weyl tensor. 

Since $m_G$ coincides with the ADM mass at spatial infinity limit, we have \footnote{See Ref. \cite{Shiromizu2023} for 
four dimensions.}
\begin{eqnarray}
m_{\rm ADM} & = &  \frac{1}{8\pi (n-3)}\int_{S_\infty} 
r \left[r^aD_ak+\frac{1}{n-2}k^2+\frac{1}{2}{}^{(n-1)}R+\frac{1}{2}\tilde k_{ab} \tilde k^{ab}+\varphi^{-2}({\cal D}\varphi)^2   \right]dA \nonumber \\
 & = &  \frac{1}{8\pi (n-3)}\int_{S_\infty} r
\left(r^aD_ak+\frac{1}{n-2}k^2  \right)dA, 
\end{eqnarray}
where, in the second equality, we used the asymptotic behavior of geometry, ${}^{(n-1)}R=O(1/r^n)$, $\tilde k_{ab}=O(1/r^{n-2})$ 
and ${\cal D}_a \varphi=O(1/r^{n-2})$. 
This expression implies that the existence of a sequence of AGPSs near the spatial infinity leads to the positivity of the ADM mass. 
Conversely, the positivity mass theorem \cite{SY1981, Witten1981} tells us the non-negativity of the ADM mass and it guarantees
the existence of the sequences of surface $S$ near the spatial infinity satisfying 
\begin{eqnarray}
\int_{S}r \left(r^aD_ak+\frac{1}{n-2}k^2  \right)dA \ge 0.
\end{eqnarray}
Note that the existence of surface $S$ is seen in an indirect way without using the asymptotic behavior of the metric \footnote{The trick for this is 
in the proof of the positive mass theorem where the asymptotic behavior is definitely crucial role.}. 

Next, we explicitly write down a new expression of the Geroch/Hawking mass in terms of the Weyl tensor. 
From the definition of the Weyl tensor, we have 
\begin{eqnarray}
R_{abcd}n^an^cr^br^d=E_{ab}r^ar^b+\frac{1}{n-2}R_{ab}(n^an^b-r^ar^b)+\frac{1}{(n-2)(n-1)}R,\label{weyl}
\end{eqnarray}
where $E_{ab}=C_{acbd}n^cn^d$ and $C_{abcd}$ is the $n$-dimensional Weyl tensor. 
The Gauss equation gives us 
\begin{eqnarray}
{}^{(n-1)}R_{ab}r^ar^b=R_{ab}r^ar^b+R_{abcd}r^ar^cn^bn^d-(K_{ab}K-K_{ac}K^c_b)r^ar^b.\label{gauss2}
\end{eqnarray}
Since 
\begin{eqnarray}
{}^{(n-1)}R_{ab}r^ar^b=-r^aD_ak-k_{ab}k^{ab}-\varphi^{-1}{\cal D}^2 \varphi
\end{eqnarray}
holds, equations (\ref{weyl}) and (\ref{gauss2}) imply us 
\begin{eqnarray}
r^aD_ak+\frac{1}{n-2}k^2 & = & -E_{ab}r^ar^b- \tilde k_{ab}\tilde k^{ab}+K_{(r)}\kappa-v_av^a-\varphi^{-1}{\cal D}^2 \varphi
\nonumber  \\
& & -\frac{n-3}{n-2}R_{ab}r^ar^b-\frac{1}{n-2}R_{ab}n^an^b
-\frac{1}{(n-2)(n-1)}R. \label{rDkE}
\end{eqnarray}
Using this, Eqs. (\ref{geroch2}) and (\ref{hawking2}) become
\begin{eqnarray}
m_G(S) & = & \frac{A_{n-2}^{\frac{1}{n-2}}}{8\pi (n-3)\Omega_{n-2}^{\frac{1}{n-2}}}\int_{S} 
\Bigl[-E_{ab}r^ar^b+\frac{1}{2}\tilde \kappa_{ab} \tilde \kappa^{ab}-\frac{1}{2} \tilde k_{ab} \tilde k^{ab}
-\frac{1}{2}\frac{n-3}{n-2}\kappa^2 \nonumber \\
& &~~~~~~~~~~~~~~~~~~ +\frac{n-3}{n-2}R_{ab}(n^an^b-r^ar^b)+\frac{n(n-3)}{2(n-2)(n-1)}R \Bigr]dA \label{geroch3}
\end{eqnarray}
and 
\begin{eqnarray}
m_H(S)& = &  \frac{A_{n-2}^{\frac{1}{n-2}}}{8\pi (n-3)\Omega_{n-2}^{\frac{1}{n-2}}}\int_{S} 
\Bigl[-E_{ab}r^ar^b+\frac{1}{2}\tilde \kappa_{ab} \tilde \kappa^{ab}-\frac{1}{2} \tilde k_{ab} \tilde k^{ab} \nonumber \\
& &~~~~~~~~~~~~~~~~~~ +\frac{n-3}{n-2}R_{ab}(n^an^b-r^ar^b)+\frac{n(n-3)}{2(n-2)(n-1)}R \Bigr]dA,
\end{eqnarray}
respectively. 
At the spatial infinity limit, one can confirm the ADM mass expression in terms of the Weyl tensor shown in Ref.~\cite{Ashtekar1984}
\begin{eqnarray}
m_{\rm ADM} = - \frac{1}{8\pi (n-3)}\int_{S_\infty} 
rE_{ab}r^ar^b dA. \label{hawking3}
\end{eqnarray}
In the quasi-local mass expressions of Eqs. (\ref{geroch3}) and (\ref{hawking3}), 
however, there are additional terms which vanish for vacuum, static and spherically symmetric region. 

Note that the left-hand side of Eq. (\ref{rDkE}) appeared in the integrands of Eqs. (\ref{geroch2}) and (\ref{hawking2}). 
The condition (\ref{defagps}) imposed in the definition of the AGPS controls the signature of this combination, that is, 
not only $E_{ab}r^ar^b$, but also non-trivial terms composed of the $n$-dimensional Ricci tensor etc through Eq. (\ref{rDkE}).

\section{Summary}

In a summary, we could see that the issue of the existence of the AGPS is strongly related to the positivity of the 
Geroch/Hawking quasi-local mass. Motivated by the AGPS condition, a new expression for the ADM mass was also presented. 
As the ADM mass formula with the Weyl tensor, we had a similar formula for the Geroch/Hawking mass.

\ack

T. S. and K. I. are supported by Grant-Aid for Scientific Research from Ministry of Education, Science, 
Sports and Culture of Japan (JP21H05182) and JSPS(No. JP21H05189). 
K. I. is also supported by JSPS Grants-in-Aid for Scientific Research (B) (JP20H01902) and JSPS Bilateral Joint 
Research Projects (JSPS-DST collaboration) (JPJSBP120227705). T. S. is also supported by JSPS Grants-in-Aid for 
Scientific Research (C) (JP21K03551). 
The authors thank the Yukawa Institute for Theoretical Physics at Kyoto University, where this work was initiated 
during the YITP-T-23-01 on "Quantum Information, Quantum Matter and Quantum Gravity". 

\appendix

\section{Quick check of Eq. (\ref{ADM-ricci})}
In $n$-dimensional spacetimes, the ADM mass is defined as the surface integral 
\begin{eqnarray}
m_{\rm ADM}=\frac{1}{16\pi}\int_{S_\infty}(\partial_i \delta g_{ij}-\partial_j \delta g_{ii} )r^jdA,
\label{A1}
\end{eqnarray}
where $\delta g_{ij}$ is the deviation from the flat metric $\delta_{ij}$, $r^i$ is the outward radial 
unit vector to $S_\infty$ and indices $i,j$ stand for the components in the Cartesian coordinate. 
Since $\delta g_{ij} \simeq \Phi \delta_{ij}$ and $\Phi \propto 1/r^{n-3}$ near the spatial infinity,
the ADM mass becomes 
\begin{eqnarray}
m_{\rm ADM}=\frac{n-2}{16\pi}\int_{S_\infty}\partial_i \Phi r^idA =\frac{(n-2)(n-3)}{16\pi}\int_{S_\infty} \frac{\Phi}{r} dA.
\end{eqnarray}
On the other hand, near the spatial infinity, we have 
\begin{eqnarray}
{}^{(n-1)}R_{ab}r^ar^b \simeq -\frac{(n-3)^2(n-2)}{2}\frac{\Phi}{r^2}. 
\end{eqnarray}
Then, we can see that Eq. (\ref{A1}) coincides with Eq. (\ref{ADM-ricci}). For the general argument, see Appendix in Ref. \cite{Tanabe2010}.

\section{Refined version of AGPS}

The refined version of AGPS is defined in Ref.~\cite{Izumi2023}. 
If a surface $S_\alpha$ satisfies 
\begin{eqnarray}
{}^{(n-2)}R -2 k {\cal D}^2 k^{-1} \ge \left( 2\alpha + \frac{n-1}{n-2}\right) k^2,
\end{eqnarray}
and $k>0$, $S_\alpha$ is called a refined AGPS. 
The advantage of the refined version is that 
it is written only with the mean curvature $k$ and the geometry of $S_\alpha$, 
while the original AGPS requires the normal derivative of $k$ to the surface. 
Therefore, in the definition of the refined AGPS, only the information of embedding is required.

The positivity of the Geroch mass $m_G$ of refined AGPSs can be seen straightforwardly 
because one can show the  positivity of the integral in Eq. \eqref{geroch1}  as
\begin{eqnarray}
\int_{S} \left({}^{(n-2)}R-\frac{n-3}{n-2} k^2 \right)dA
&\ge& 2 \int_{S} \left[ k^{-2} \left({\cal D} k\right)^2 +\left(\alpha + \frac{2}{n-2} \right) k^2 \right]dA
\nn
&>&0.
\end{eqnarray}
Furthermore, $m_H\ge m_G$ results in the positivity of the Hawking mass. 

It would be interesting to investigate another quasi-local mass, the Bartnik mass~\cite{Bartnik:1989zz}.
Suppose an induced metric $h_{ab}$ and an extrinsic curvature $k_{ab}$ are given for $S$.
Then, the definition of the Bartnik mass $m_B$ is 
\begin{eqnarray}
m_B(S) = \inf_{\tilde M}  m_{ADM}(\tilde M),
\end{eqnarray}
where the infimum is taken for any asymptotically flat $(n-1)$ dimensional space without minimal surfaces and with the non-negative scalar curvature
whose boundaries are composed only of an infinity and $S$ with the given $h_{ab}$ and $k_{ab}$, 
that is, $\tilde M$ is any admissible extension from $S$ to the asymptotically flat infinity. 
In asymptotically flat spaces with the nonnegative scalar curvature, the area of a refined AGPS is shown to be bounded as~\cite{Izumi2023}
\footnote{
The ADM mass in this paper differs by a factor from that in Ref.~\cite{Izumi2023} which follows the definition in Ref.~\cite{Bray:2007opu}.
}
\begin{eqnarray}
A \le \Omega_{n-2} \left[ \frac{(n-1)+2(n-2)\alpha}{1+(n-2)\alpha} \frac{8 \pi m_{ADM}}{(n-2)\Omega_{n-2}}  \right]^{\frac{n-2}{n-3}}.
\end{eqnarray}
Going back to the proof of the higher dimensional Penrose inequality~\cite{Bray:2007opu}, and following the discussion in Ref.~\cite{Izumi2023}, 
we can see that this areal inequality is written as a lower bound for the mass
\begin{eqnarray}
m_{ADM} \ge \frac{(n-2) \Omega_{n-2}}{8 \pi} \frac{1+(n-2)\alpha}{(n-1)+(n-2)\alpha} \left( \frac{A}{\Omega_{n-2}}\right)^{\frac{n-3}{n-2}}.
\end{eqnarray}
Since this inequality is satisfied in any possible extended space $\tilde M$ from $S$ to the asymptotically flat infinity, 
it shows the lower bound of the Bartnik mass and its positivity.



\begin{thebibliography}{9}

\bibitem{Shiromizu2017}
T.~Shiromizu, Y.~Tomikawa, K.~Izumi and H.~Yoshino,
PTEP \textbf{2017}, no.3, 033E01 (2017).

\bibitem{Izumi2021}
K.~Izumi, Y.~Tomikawa, T.~Shiromizu and H.~Yoshino,
PTEP \textbf{2021}, no.8, 083E02 (2021).

\bibitem{Izumi2023}
K.~Izumi, Y.~Tomikawa, T.~Shiromizu and H.~Yoshino,
PTEP \textbf{2023}, no.4, 043E01 (2023).

\bibitem{Penrose}
R.~Penrose, Annals N. Y. Acad. Sci. \textbf{224}, 125 (1973);\\
P.~S.~Jang and R.~M.~Wald, J. Math. Phys. \textbf{18}, 41 (1977);\\
G. Huisken and T. Ilmanen, J. Diff. Geom. {\bf 59}, 353 (2001);\\
H. Bray, 
J. Differential Geom., \textbf{59}
(2001), 177-267. 

\bibitem{Geroch1973}
R.~Geroch, Ann. N. Y. Acd. Sci. {\bf 224}, 108 (1973). 

\bibitem{Hawking1968}
S.~Hawking,
J. Math. Phys. \textbf{9}, 598-604 (1968).

\bibitem{Ashtekar1984}
A. Ashtekar and A. Magnon, J. Math. Phys. \textbf{25}, 2682 (1984).

\bibitem{Tanabe2010}
K.~Tanabe, N.~Tanahashi and T.~Shiromizu,
J. Math. Phys. \textbf{50}, 072502 (2009).

\bibitem{Shiromizu2023}
T.~Shiromizu and K.~Izumi,
PTEP \textbf{2023}, no.10, 103E02 (2023).

\bibitem{Shiromizu2010}
R.~Mizuno, S.~Ohashi and T.~Shiromizu,
Phys. Rev. D \textbf{81}, 044030 (2010).




\bibitem{SY1981}
R.~Schoen and S.~T.~Yau,
Commun. Math. Phys. \textbf{79}, 231 (1981). 

\bibitem{Witten1981}
E.~Witten,
Commun. Math. Phys. \textbf{80}, 381 (1981).

\bibitem{Bartnik:1989zz}
R.~Bartnik,
Phys. Rev. Lett. \textbf{62}, 2346-2348 (1989);\\
R.~Bartnik,
Energy in General Relativity, Tsing Hua Lectures on Geometry and Analysis, International Press of Boston, Cambridge MA, 5-27, 1997. 

\bibitem{Bray:2007opu}
H.~L.~Bray and D.~A.~Lee,
Duke Math. J. \textbf{148}, 81-106 (2009).
\end{thebibliography}
\end{document}